\documentclass[fleqn,10pt,twocolumn]{wlscirep}
\usepackage[utf8]{inputenc}
\usepackage[T1]{fontenc}
\usepackage{bm}

\usepackage{siunitx}
\usepackage{color}
\usepackage{color}
\usepackage{xcolor}
\usepackage{hyperref}
\usepackage{comment}

\usepackage{graphicx}
\usepackage{framed}
\usepackage{esvect}
\usepackage{amsmath,amssymb,color,bm}
\usepackage{mathrsfs,dsfont}
\usepackage[dvipsnames]{xcolor}
\usepackage{comment}
\usepackage{multirow}
\usepackage{framed}
\usepackage{color}

\newcommand{\costh}{\ensuremath{\langle \cos^2 \theta_{\text{2D}} \rangle}}

\newcommand{\CSS}[0]{$\mathrm{CS_2}$}
\newcommand{\ECFG}[0]{\ensuremath{\vec{\mathbf{{E}}}_\mathrm{CFG}}}
\newcommand{\fcfg}[0]{\ensuremath{f_\mathrm{CFG}}}

\newcommand\tb[1]{\textbf{#1}}

\newcommand\beq{\begin{equation}}
\newcommand\eeq{\end{equation}}
\newcommand\beqa{\begin{eqnarray}}
\newcommand\eeqa{\end{eqnarray}}

\newcommand\re[1]{\textnormal{Re}\left[#1\right]}

\def\v{\tb{v}}

\def\.{\cdot}
\def\1{^{-1}}
\def\2{^{-2}}
\def\3{^{-3}}

\AtBeginDocument{\renewcommand{\hbar}{{\mkern0.75mu\mathchar'26\mkern-6.75mu h}}}

\title{Optical centrifuge as a probe of strong dissipative coupling between a molecular rotor and superfluid helium}

\author[1]{Ian MacPhail-Bartley}
\author[1]{S{\"o}ren E. Mahr}
\author[1]{Cameron E. Peters}

\author[2]{Baptiste Coquinot}
\author[2]{Volker Karle}
\author[3]{Giacomo Bighin}
\author[2]{Ragheed Alhyder}
\author[2]{Mikhail Lemeshko}

\author[4]{Henrik Stapelfeldt}
\author[1,*]{Valery Milner}

\affil[1]{Department of Physics and Astronomy, The University of British Columbia, Vancouver, British Columbia, Canada}
\affil[2]{Institute of Science and Technology Austria (ISTA), Klosterneuburg, Austria}
\affil[3]{Department of Physics, University of Zagreb, Zagreb, Croatia}
\affil[4]{Department of Chemistry, Aarhus University, Aarhus, Denmark}

\affil[*]{e-mail: vmilner@phas.ubc.ca}

\begin{abstract}
A macroscopic manifestation of superfluidity is that objects moving through liquid helium experience negligible friction below the Landau critical velocity. How this frictionless motion breaks down at the nanoscale remains an open question. Molecules embedded in helium nanodroplets represent a well-controlled system for studying this breakdown, yet none has reached the regime of strong dissipative coupling, when energy transfer from the molecule to the superfluid dominates the observed dynamics. Molecular rotation, induced by short laser pulses, offer a suitable probe to reach rotational energies in the range of the roton gap, where superfluid helium supports a large number of elementary excitations. However, the solvation shell around a rotating molecule caps the energy reachable by a free rotor after impulsive excitation well below the roton excitation energy. Here we show that continuous driving with an ultraslow optical centrifuge overcomes this limitation: the strong field dresses the molecule into pendular states whose energies fall within the spectrum of the collective excitations of the superfluid, placing the system in the strong-dissipation regime. The resulting rapid thermalization locks the molecule to the rotating field until the rotation-induced level splittings overtake the thermalization rate, beyond which the molecular alignment is progressively lost. Our approach offers a direct measurement of the molecule-bath coupling in a quantum fluid.
\end{abstract}

\begin{document}

\flushbottom
\maketitle
\thispagestyle{empty}

Superfluid helium supports frictionless motion of macroscopic objects, provided their velocity remains below a critical value, above which the moving object dissipates energy by exciting roton pairs~\cite{Allum1977} and vortex rings~\cite{Nancolas1985} in the fluid. The Landau critical velocity $v_{\mathrm{L}}$~\cite{Landau1941}, at which roton creation is possible, gives an upper bound. Probing the onset of this dissipation on the atomic scale requires a microscopic object whose motion can be both controlled and measured. Early experiments accelerated ions through bulk liquid helium in an external electric field, and revealed dissipation consistent with the scattering of ions from the collective excitations of the superfluid~\cite{Reif1960}. However, an ion in liquid helium does not move as a bare particle: depending on its charge state, it forms either a ``snowball'' of solidified helium or a cavity-like ``bubble''~\cite{Atkins1959}, whose effective hydrodynamic mass is substantially modified, often depending on the pressure and temperature of the liquid~\cite{Gunther1996}. Furthermore, since the ion continuously accelerates in the applied field, direct control over its velocity is limited.

Helium nanodroplets have opened a complementary route to the same physics~\cite{Toennies2004}. For instance, solvated atoms and molecules can be repelled from the droplet by electronic excitation, because the dopant-helium interaction becomes repulsive. It was found that the terminal velocities of the ejected atoms and molecules were close to $v_{\mathrm{L}}$ despite the deposited energy into the droplet by the dopants~\cite{Brauer2013}. Here, a bubble again forms around the electronically excited atom or molecule~\cite{Federmann1999}, potentially modifying the coupling to the superfluid~\cite{Milner2023}. In parallel with these dynamical studies of moving objects, high-resolution IR spectroscopy has revealed signatures of microscopic superfluidity in the frequency domain: molecules embedded in helium droplets exhibit sharp, well-resolved rotational lines, demonstrating that they rotate freely inside the quantum fluid~\cite{Grebenev1998}. Rotational line broadening has been found to increase as the rotational energy approaches the roton gap, further indicating that collective excitations underlie the dissipation mechanism~\cite{Choi2006}.

Ultrafast laser pulses offer a dynamical probe of both the internal degrees of freedom, and the translational motion~\cite{Christensen2025}. The dynamics of impulsively excited vibrational wave packets in molecules embedded in helium droplets~\cite{Schlesinger2010} were well reproduced by a dissipative model in which the friction vanishes once the velocity of the atoms relative to the droplet falls below the critical velocity. However, the control over this velocity is very limited, constrained by the discrete set of available vibrational levels. Rotational wave packets, produced by femtosecond ``kick'' pulses, give access to the much denser rotational spectrum of molecules embedded inside~\cite{Pentlehner2013} and on the surface~\cite{Kristensen2025} of helium nanodroplets. Impulsive rotational excitation has also enabled the study of energy levels inaccessible to IR spectroscopy, which can only probe the lowest lying states that are populated at the 0.37~K temperature of the droplets. Yet it faces an obstacle of its own: the significantly increased molecular centrifugal distortion constant due to the solvation shell of helium atoms puts an upper limit, hereafter referred to as the ``centrifugal wall'', on the accessible rotational levels~\cite{Chatterley2020,Cherepanov2021}, well below the roton gap (with an exception of very light molecules \cite{Qiang2022}).

Common to all these approaches is the inability to bring the molecule-superfluid system into the regime of strong dissipative coupling, in which the energy scale of the rotor matches that of the quasiparticles of the fluid. The coupling strength could not be controlled: an ejected Rydberg species accelerates only up to $v_{\mathrm{L}}$, whereas the rotational frequencies of a kicked molecule cannot approach the critical scale owing to the centrifugal wall, confining the system to the perturbative domain of angulon theory~\cite{Schmidt2015,Lemeshko2017}. Strong optical driving changes this situation by introducing a new energy scale -- that of the pendular states of the molecule trapped in the rotating field. For the field strengths employed here, the pendular energies fall within the spectrum of the collective excitations of the superfluid, bringing the molecule-bath system into the regime of rapid thermalization. Forced rotation in an optical centrifuge then offers a quantitative handle on this dissipative coupling: the bath continuously relaxes the molecule towards the bottom of the rotating trap, but only as long as the thermalization outpaces the rotation. The frequency at which the driven molecule loses its alignment to the centrifuge field therefore provides a direct measure of the molecule-specific thermalization rate, and thus its coupling to the superfluid.

\begin{figure}
\includegraphics[width=1\columnwidth]{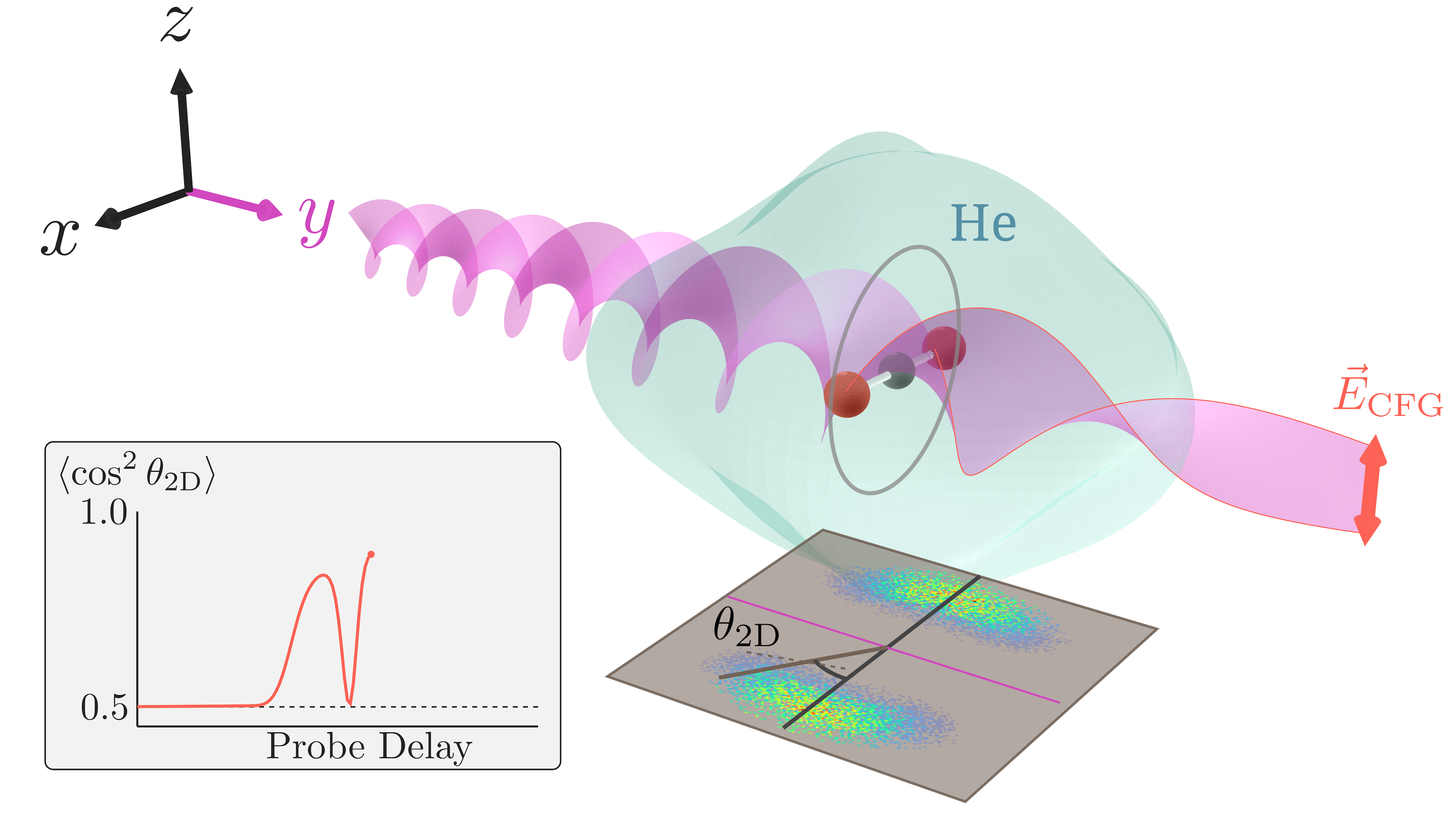}
\caption{\label{fig:Geometry} Geometry of the Velocity Map Imaging experiment. The magenta corkscrew represents the electric field of the optical centrifuge (\ECFG{}), propagating to the right along the $y$ axis. The VMI spectrometer, oriented along the $z$ axis, projects the ion fragment velocities onto the $xy$ plane of the detector. $\theta_{\text{2D}}$ is the angle of a detected ion fragment, measured with respect to the $xz$ plane of centrifuge rotation. A carbon disulfide (\CSS{}) molecule inside a helium droplet is shown aligned to the centrifuge, with a simulated distribution of S$^+$ fragments shown below. The inset illustrates $\costh{}(t)$ for a molecule following the rotation of \ECFG{} during the initial portion of the centrifuge pulse.}
\end{figure}
An optical centrifuge (illustrated in Fig.~\ref{fig:Geometry}) is a linearly polarized laser pulse whose polarization vector undergoes controlled rotation about the propagation axis~\cite{Karczmarek1999, Villeneuve2000}. Through the interaction with the induced dipole, the field exerts a torque on the molecule, forcing it to follow the rotating polarization with angular accelerations of up to 100~GHz/ps~(Ref.~\citenum{Macphail2020}). Attempts to spin molecules inside helium nanodroplets with centrifuges designed for the gas phase proved unsuccessful~\cite{Jorgensen2018, Macphail2024}, owing to the higher effective moments of inertia of a solvated molecule requiring lower angular accelerations than those probed in the gas phase. We recently demonstrated that molecules in helium droplets can be centrifuged in the limiting case of zero angular acceleration, i.e. at a constant rotational frequency~\cite{Macphail2026}. However, the limited phase stability of that centrifuge prevented the detection of molecular rotation at frequencies above the centrifugal wall. Here, we employ the recently developed ``ultraslow'' optical centrifuge~\cite{Wang2026} (usCFG, see Supplementary Information), capable of arbitrarily low angular acceleration and offering substantially improved phase stability~\cite{Mahr2026}, to spin \CSS{} and OCS molecules embedded in helium nanodroplets well past their respective centrifugal walls, significantly faster than previously achieved with a single linearly polarized pulse.

To detect molecular rotation, we Coulomb-explode the molecules with an intense femtosecond probe pulse and image the recoiling ionic fragments using Velocity Map Imaging (VMI, see Supplementary Information). We define $\theta_\mathrm{2D}$ as the angle between the detected ion fragment, which for axial recoil coincides with the projection of the molecular axis onto the detector plane ($xy$), and the plane of centrifuge rotation ($xz$), as shown in Fig.~\ref{fig:Geometry}. The degree of molecular alignment is quantified by \costh{}, averaged over the molecular ensemble. If the molecules follow the rotating polarization of the field, $\costh(t)$ oscillates at twice the rotational frequency of the centrifuge, passing through a maximum every time the molecular axis crosses the detector plane, as schematically depicted in the inset of Fig.~\ref{fig:Geometry}. Throughout the paper, we quote frequencies as measured, i.e. in terms of the signal oscillation frequency, unless explicitly stated otherwise.

\begin{figure*}
\centering
\includegraphics[width=0.95\textwidth]{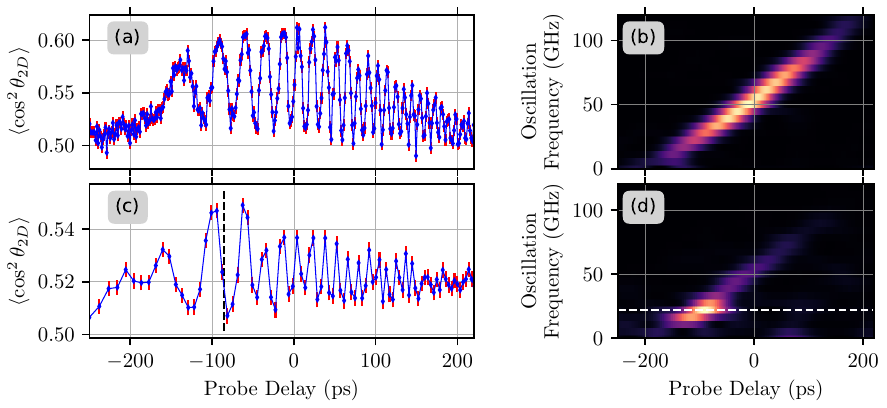}
\caption{\label{fig:gas_comparison} Degree of alignment in the ensemble of centrifuged \CSS{} molecules, expressed as $\costh(t)$ (blue markers with red error bars) and measured in (\textbf{a}) the seeded molecular jet, and (\textbf{c}) helium nanodroplets. Blue lines connect the data points to help guide the eye. Applying a short-time Fourier transform to both signals results in the corresponding two-dimensional spectrograms, shown in panels (\textbf{b}) and (\textbf{d}). The white dashed line in panel (\textbf{d}) marks the ``centrifugal wall'' of the \CSS{}-helium complex, occurring around \SI{22}{\giga \hertz}~(Ref.~\citenum{Cherepanov2021}).}
\end{figure*}

Fig.~\ref{fig:gas_comparison}(\textbf{a}) shows the oscillatory $\costh(t)$ signal from centrifuged \CSS{} molecules seeded in a supersonic jet of helium gas. The gradual decrease of the oscillation period indicates that the molecules follow the rotating polarization of the centrifuge field, accelerating from rest at \SI{-200}{\pico\second} to beyond \SI{100}{\giga\hertz} by the end of the pulse. The envelope of the oscillations traces the approximately Gaussian intensity profile of the laser pulse (\SI{320}{\pico\second} full width at half maximum, FWHM, centered around $\SI{0}{\pico\second}$ delay), since the degree of alignment to \ECFG{} strongly depends on the field intensity~\cite{Larsen1999, Stapelfeldt2003}. The relatively low degree of alignment after the end of the usCFG pulse reflects the fact that the rotational energy transferred to the molecules by the centrifuge is comparable to the initial thermal rotational energy in the jet ($T_\mathrm{jet}\sim\SI{10}{\kelvin}$).

To visualize the constant angular acceleration of the molecular spinning, we apply a short-time Fourier transform to the experimental signal in a sliding time window, and plot the calculated power spectra as a function of time. The resulting two-dimensional spectrogram, shown in panel (\textbf{b}), exhibits a single spectral component, ascending linearly with time -- a direct signature of the constant angular acceleration of the driven molecules.

The rotational dynamics of \CSS{} molecules embedded in helium droplets and driven by the same centrifuge pulse as in Fig.~\ref{fig:gas_comparison}(\textbf{a}) are presented in panels (\textbf{c}) and (\textbf{d}). The oscillations in panel (\textbf{c}) persist over a similar timescale, and the spectrogram in panel (\textbf{d}) traces the driven rotation up to \SI{80}{\giga\hertz} -- well beyond the centrifugal wall of \SI{22}{\giga\hertz} (horizontal dashed line), previously established in femtosecond-kick experiments~\cite{Cherepanov2021}. The helium environment results in a lower amplitude of the \costh{} oscillations, owing to the non-axial recoil of the S$^+$ ions Coulomb-exploded inside the droplets~\cite{Christensen2016,Chatterley2020}.

\begin{figure}[t]
\includegraphics[width=1\columnwidth]{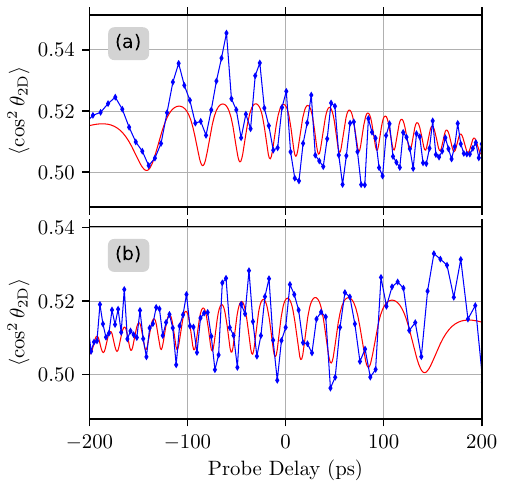}
\caption{\label{fig:CSS_directionality}
The alignment dynamics of \CSS{} molecules embedded in helium droplets spun by accelerating and decelerating ultraslow optical centrifuges, with $\costh(t)$ shown in (\textbf{a}) and (\textbf{b}), respectively. The experimental data (connected blue diamonds) is compared with the results of our numerical simulations (red traces).}
\end{figure}
To understand the observed dynamics, we developed a theoretical model describing the interaction of a molecule, continuously driven by the strong laser field of the centrifuge, with the many-body helium environment. We work in the frame co-rotating with the centrifuge at frequency $\Omega_0(t)$, where the field polarization is stationary, and assume that the bath dynamics are fast on the timescale of the driven molecular rotation. The molecular Hamiltonian reads
\begin{equation}
  H_{\rm rot}^{\rm RF}(t) = \hbar B\,\mathbf{J}^2 - \hbar \Omega_0(t)\,J_z
  - \frac{\Delta\alpha}{4}\,|\textbf{E}_{\rm CFG}(t)|^2 \,\sin^2\!\theta\cos^2\!\phi,
  \label{eq:Hrot}
\end{equation}
where $B$ is the gas-phase rotational constant of the molecule, $\Delta\alpha{=}\alpha_\parallel{-}\alpha_\perp$ is the polarizability anisotropy, and $\textbf{E}_{\rm CFG}(t)$ is the laser field amplitude. Angles $\theta$ and $\phi$ specify the orientation of the molecular axis in the rotating frame: $\theta$ is measured from the laser propagation direction ($y$ in Fig.~\ref{fig:Geometry}) and $\phi$ from \ECFG{} within the plane of rotation ($xz$). Simulations of the dynamics of centrifuged molecules in the gas phase, using Hamiltonian~(\ref{eq:Hrot}), are in good agreement with our experimental observations in the molecular jet.

In contrast to the previously used solvation-shell description of freely rotating molecules with renormalized rotational and centrifugal-distortion constants, here we treat the molecule as a driven rotor, and account for the helium environment through a phenomenological thermalization rate. We numerically diagonalize $H_{\rm rot}^{\rm RF}(t)$ at every time step of the simulation, each corresponding to a different rotational frequency. The slow variation of $\Omega_0(t)$ justifies the use of the instantaneous Hamiltonian, while the fast thermalization by the bath erases the memory between successive time steps, allowing us to treat them independently. This approximation was validated by computing the full Lindbladian dynamics (see Supplementary Information).

As long as the energy of the rotor, determined by the instantaneous frequency of the centrifuge, remains below the depth of the centrifuge potential well, dictated by its instantaneous intensity, i.e. $\Omega_0 \lesssim \Omega_\mathrm{loc}$, where $\hbar\Omega_\mathrm{loc} = \sqrt{\hbar B\Delta\alpha |\textbf{E}_{\rm CFG}|^2}$ is the pendular frequency of the trap, the ground state of $H_{\rm rot}^{\rm RF}$ remains angularly localized along \ECFG{}, co-rotating with the centrifuge. Because of the assumed fast dynamics of the bath compared to the molecule's rotation, we compute the corresponding self-energy perturbatively~\cite{Lemeshko2017}, arriving at only a weak renormalization of the energy eigenstates (see Supplementary Information). The much stronger effect of the helium environment is dissipative. At each instant, the bath thermalizes the ensemble by driving the molecular density matrix towards the corresponding thermal distribution in the lab frame -- that is, towards alignment with the instantaneous direction of \ECFG{}.

This process is modeled through a phenomenological master equation, which yields a steady state density matrix $\rho_{\rm ss}^{\rm RF}$. In the basis of the instantaneous rotating Hamiltonian $H_{\rm rot}^{\rm RF}|n\rangle=\tilde{E}_n|n\rangle$, the steady state reads:
\begin{equation}
  \left(\rho_{\rm ss}^{\rm RF}\right)_{mn}
  =
  \frac{
    \langle m|\,
    e^{-\left(\hbar B\mathbf{J}^2 - \frac{\Delta\alpha}{4} |\textbf{E}_{\rm CFG}|^2\,\sin^2\!\theta\cos^2\!\phi\right)/k_{\rm B}T}
    |n\rangle / Z
  }{1 + i\tau\left(\tilde{E}_m - \tilde{E}_n\right)/\hbar},
  \label{eq:rho_ss}
\end{equation}
\noindent where $\tau$ is the thermalization time constant, assumed uniform for all states, $T$ is the droplet temperature, and $Z$ is the partition function. As demonstrated below by comparison with the experiment, Eq.~\eqref{eq:rho_ss} captures the observed dynamics. At low rotational frequencies and short thermalization times, the denominator is close to unity, and the steady state reduces to the pendular state of the stationary trap, aligned to \ECFG{} and continuously restored by the bath. This aligned state, however, is not an eigenstate of the rotating-frame Hamiltonian but a coherent superposition of its eigenstates. As $\Omega_0$ grows, the level splittings $\tilde{E}_m - \tilde{E}_n$ in the rotating frame increase and, as they reach $\hbar/\tau$, suppress the corresponding coherences, thus dephasing the superposition and lowering the degree of alignment to \ECFG{}.

\begin{figure}[t]
\includegraphics[width=1\columnwidth]{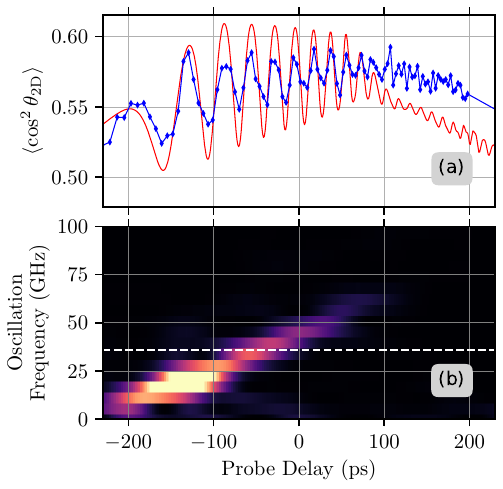}
\caption{\label{fig:OCS}
(\textbf{a}) The alignment dynamics of centrifuged OCS molecules embedded in helium droplets. The experimental data (connected blue diamonds) is compared with the results of our numerical simulations (red traces). (\textbf{b}) Short-time Fourier transform, with the ``centrifugal wall' indicated by the white dashed line.}
\end{figure}
The results of our simulations are shown with a red curve in Fig.~\ref{fig:CSS_directionality}(\textbf{a}). The best agreement with the experimental data (blue diamonds) is obtained with the peak intensity of the centrifuge field $I=\frac{1}{2}c\epsilon_0|\textbf{E}_{\rm CFG}|^2= \SI[parse-numbers = false]{3\cdot 10^{10}}{W/cm^2}$ and the effective thermalization time constant $\tau=\SI[parse-numbers = false]{10}{ps}$ (at high rotational frequency). Using the gas-phase molecular constants of \CSS{}, the rotating potential well can sustain a localized state only up to a rotational frequency of $\Omega_\mathrm{loc}/2\pi \approx \SI[parse-numbers = false]{30}{GHz}$ ($\SI{60}{GHz}$ oscillation frequency). The observed rotation persists beyond this threshold because the relevant level splittings satisfy $\tau(\tilde{E}_m-\tilde{E}_n)/\hbar \sim 1$, meaning that the bath restores the aligned superposition faster than the rotation-induced effects destroy it. As the splittings grow with $\Omega_0$, the dephasing progressively prevails, delocalizing the angular distribution and causing the gradual decrease of the observed alignment. The amplitude of the simulated \costh{} oscillations has been scaled by a constant factor to account for the non-axial recoil of the fragments~\cite{Chatterley2020}.

To further validate our theoretical model, we note that since it assumes no memory between consecutive time steps, its predictions at any instant depend only on the instantaneous frequency and intensity of the centrifuge, and not on the history of the sweep. To test this assumption, we modified the usCFG in such a way as to make it \textit{decelerate} from a high starting frequency to zero. As shown in Fig.~\ref{fig:CSS_directionality}(\textbf{b}), both the experimental and the numerical results reproduce the accelerating case of panel (\textbf{a}) in reverse. Here, as the centrifuge frequency decreases, the level splittings drop with it, allowing the superfluid bath to restore the angular localization, with a correspondingly growing oscillation amplitude. This behavior stands in sharp contrast to the conventional centrifuge action in the gas phase, which relies on the molecule being captured at low frequency and adiabatically carried upward. The observed reversibility underscores the dominant role of the superfluid bath in the regime  of strong dissipative coupling explored in this work.

To investigate how the observed dynamics depend on the interaction between a specific molecule and the superfluid environment, we repeated the experiment with carbonyl sulfide (OCS). In the gas phase, OCS responds to the centrifuge drive similarly to \CSS{}~(Ref.~\citenum{Macphail2020}), while its coupling to the helium bath, estimated with density functional theory, is approximately three times weaker~\cite{Higgins1999, Farrokhpour2013}. Within our model, the weaker coupling implies a longer thermalization time $\tau$ and, consequently, a lower rotational frequency at which the dephasing prevails and the alignment to \ECFG{} is lost. This is indeed confirmed by the experimentally recorded \costh{} for the centrifuged OCS in helium droplets, shown in Fig.~\ref{fig:OCS}. We believe that the overall higher amplitude of the oscillations, as compared with \CSS{}, reflects better axial recoil of the fragments in the Coulomb explosion of OCS, and bears no relation to the rotational dynamics.

From the short-time Fourier analysis, plotted in panel (\textbf{b}) of Fig.~\ref{fig:OCS}, we note that OCS molecules spin beyond the \SI{36}{\giga\hertz} centrifugal wall, determined in the femtosecond kick experiments~\cite{Chatterley2020}. Yet the oscillation amplitude starts declining much earlier (notably, while the laser intensity is still rising) and decreases much faster than for \CSS{}. Our numerical calculations (solid red) reproduce this behavior with a ten times longer thermalization time, $\tau = \SI{100}{\pico\second}$, consistent with the quadratic dependence of the thermalization rate on the approximately threefold weaker coupling of OCS to the bath. We further note that the oscillatory part of \costh{} is superimposed on a strong background, which follows the intensity profile of the centrifuge pulse. We interpret this background as planar alignment, akin to that induced by circularly polarized pulses in helium droplets~\cite{Pickering2018}: the molecules remain localized near the plane of usCFG rotation even after their localization within the rotating trap can no longer be restored by the bath.

\begin{figure}
\includegraphics[width=1\columnwidth]{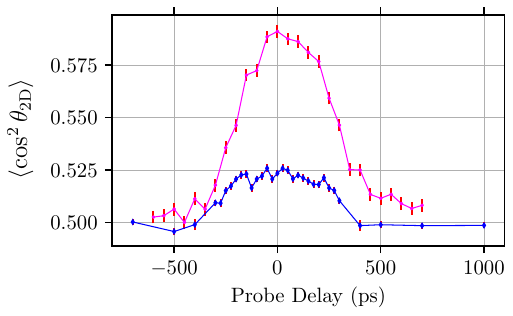}
\caption{\label{fig:LongScan}
\costh{} signal averaged over the random initial orientation of the centrifuge polarization (see Supplementary Information for details). Connected blue (magenta) diamonds with red error bars correspond to \CSS{} (OCS) molecules embedded in helium droplets.}
\end{figure}
Measuring the rate of thermalization by recording \costh{} \textit{after} the centrifuge proves challenging with our current usCFG. By the end of the pulse, the decay of molecular alignment is either too fast to be distinguished from the falling edge of the intensity profile (as in the case of \CSS{}), or the molecules have already been delocalized within the rotating trap (OCS). Moreover, the residual planar alignment of OCS is an adiabatic effect: the instantaneous degree of angular confinement follows the field intensity and vanishes with it at the end of the pulse, in contrast to the rotational excitation accumulated by a molecule following the accelerating trap. We illustrate this in Fig.~\ref{fig:LongScan} by plotting the experimental signals for both molecules, averaged over the random initial orientation of the centrifuge polarization (hence the smooth, oscillation-free profile; see Supplementary Information). In both cases, \costh{} returns to the value of 0.5 for an isotropic distribution shortly after the pulse, as expected from the above arguments.

In summary, we have developed and used an ultraslow optical centrifuge to drive the accelerated rotation of \CSS{} and OCS molecules embedded in superfluid helium nanodroplets, reaching rotational frequencies far beyond the centrifugal wall that limits impulsively kicked rotors. While the wall stems from the large centrifugal distortion constant of a helium-solvated molecule, owing to the solvation shell~\cite{Chatterley2020,Cherepanov2021}, it does not constrain a continuously driven molecule. The strong field of the centrifuge dresses it into pendular states, whose energies fall within the range of the collective excitations of the bath, placing the molecule-superfluid system in the regime of strong dissipative coupling.

Contrary to common intuition, it is not just the coherent drive of the centrifuge, but also the interaction with the droplet environment, that sustains the molecular rotation in this regime. Rapid thermalization continuously rebuilds the aligned state in the rotating trap, while the rotation-induced level splittings act to dephase it. The competition between the two converts the centrifuge into a quantitative probe of the molecule-bath coupling. The rotational frequency at which the driven molecule loses its alignment reflects the molecule-specific thermalization time, here found to differ by an order of magnitude between \CSS{} ($\tau = \SI[parse-numbers = false]{10}{\pico\second}$) and OCS ($\tau = \SI{100}{\pico\second}$), in line with the weaker coupling of the latter to the bath. The observed reversibility of the driven dynamics with respect to the sweep direction further confirms the bath-dominated mechanism of rotational excitation.

Several directions follow from this work. The assumption of a single thermalization time for all states is rather crude: infrared spectroscopy shows that the coherence lifetimes of the low-energy rotational states are longer than the thermalization times inferred here~\cite{Choi2006}, calling for a state-resolved treatment of the molecule-bath coupling. On the experimental side, an improved usCFG capable of sudden truncation at a predefined terminal frequency is being developed in our group, with its sharp falling edge enabling direct time-domain measurements of lifetimes on the \SI{10}{\pico\second} scale.

Finally, at substantially higher intensities, currently out of reach in our setup, the pendular energies would be lifted above the spectrum of the bath excitations, suppressing the coupling and restoring long-lived coherent driving. In that regime, the rotational Doppler shift of the bath excitations in the co-rotating frame is predicted to open a dissipation channel only at a well-defined driving frequency~\cite{Coquinot2026}. This will turn the crossover from slow to fast dissipation into a switchable, frequency-controlled gate. Together, these developments establish continuous driving with an optical centrifuge as a versatile platform for studying how a superfluid exchanges energy with an embedded quantum rotor.

\section*{Acknowledgments}
This research has been supported by the grants from CFI, BCKDF, and NSERC, and carried out under the auspices of the UBC Center for Research on Ultra-Cold Systems (CRUCS). We would also like to thank Jing-Lun Li and Georgios M. Koutentakis for useful discussions. HS acknowledges the support from Villum Fonden through a Villum Investigator Grant No. 25886. GB acknowledges support from the project ``Implementation of cutting-edge research and its application as part of the Scientific Center of Excellence for Quantum and Complex Systems, and Representations of Lie Algebras", Grant No. PK.1.1.10.0004, co-financed by the European Union through the European Regional Development Fund - Competitiveness and Cohesion Programme 2021-2027. BC acknowledges support from the NOMIS Foundation. RA acknowledges funding from the Austrian Academy of Science \"{O}AW grant No. PR1029OEAW03.

\newpage
\onecolumn
\begin{centering}
\textsc{\huge Supplementary Information}
\end{centering}
\section{Experimental Methods}
\subsection{The Ultraslow Optical Centrifuge}\label{usCFG}
The usCFG used in this work was described in detail in Ref.~\cite{Wang2026}. It is formed from the uncompressed output of a chirped pulse amplifier (Spectra-Physics Spitfire Ace), with a bandwidth of \SI{\approx8.2}{\nano \metre} (full width at half maximum), centred at \SI{802}{\nano \metre}. The usCFG has an approximately Gaussian intensity profile with a FWHM duration of \SI{\approx 320}{\pico \second}, and rotational frequency $\fcfg{}(t)$ accelerating from approximately 10-\SI{30}{\giga \hertz} over that time. The rotational acceleration and frequency at $t=0$ of the usCFG can be modified with negligible impact on the pulse's spatiotemporal profile. The centrifuge is interferometrically stable, meaning that the orientation of $\ECFG(t)$ (i.e. the rotational phase) is the same for every laser pulse. This means that the orientation of the centrifuge for a given probe delay is identical for each laser pulse. Active phase stabilization is possible by tracking the phase of the centrifuge spectrum and correcting for drift with a half wave plate~\cite{Mahr2026}.

\subsubsection{Degree of alignment after the optical centrifuge}
In Fig.~5, the asymptotic values of \costh{} are determined by the final superposition of rotational states after the centrifuge. Values of $\costh{} \approx 0.5$ by \SI{\approx 500}{\pico \second} imply that there is no lasting molecular rotation, and that the relaxation time is shorter than the falling edge of the centrifuge's intensity profile.

\subsubsection{Lack of oscillations present in Figure~5}
To ensure that the specific time delays chosen after the centrifuge would not impact the measured values of \costh{}, the centrifuge was passed through a half-wave plate in a continuously-rotating motorized mount. This averages the orientation of \ECFG{} at each probe delay across all angles within the plane of rotation, effectively scrambling the phase of each centrifuge pulse.

\subsection{The Helium Droplet System}\label{droplets}
The high vacuum system with the Velocity Map Imaging apparatus used to study molecules in helium nanodroplets, as well as gas phase molecules produced in molecular jets, is described in detail in Ref.~\cite{Macphail2024}, and only described here in brief detail. For a recent review of the technique, see Ref.~\cite{Slenczka2022}.

Helium droplets were formed following supersonic expansion of pre-cooled high-purity helium gas through a \SI{5}{\micro \metre} orifice into vacuum. The helium droplets pass through a \SI{1}{\milli \metre} skimmer into a subsequent chamber, where they then doped, by controlling the partial pressure of the target dopant, to contain at most one \CSS{} or OCS molecule. The droplets were produced at stagnation conditions of \SI{30}{bar} and \SI{16}{K} or \SI{18}{K} for the two respective molecules. Under these conditions, the helium droplets are expected to consist of approximately 3000-5000 He atoms.

\subsubsection{Velocity Map Imaging}
The optical centrifuge is focused with a lens ($f = \SI{20}{\cm}$) to a nominal non-ionizing peak intensity of $\approx5 \times 10^{11} \pm\SI{5E11}{\watt \per \cm^2}$ onto the doped helium droplet beam. A short (\SI{\approx120}{\femto \second}) probe pulse whose linear polarization is perpendicular to the detector, aligned collinear to the centrifuge, is focused with the same lens to a peak intensity of \SI{\approx4E14}{\watt \per \cm^2} to Coulomb explode the target molecule. Velocities of the resulting S$^+$ ion fragments from OCS and \CSS{} are measured with a Time of Flight Velocity Map Imaging spectrometer~\cite{Eppink1997}, which utilizes an electrostatic lens with a repeller voltage of \SI{6}{\kilo \volt} to project a 2D velocity map of the ions onto a dual-layer microchannel plate detector. The detector is connected to a phosphor screen, where a 2D image of the ion velocity distribution is recorded with a camera. The angle $\theta_{\mathrm{2D}}$ (see Fig~1) is evaluated for each ion hit, with \costh{} evaluated over several thousand ions. The arrival time of the probe is varied with respect to the centrifuge using a translation stage, to measure  $\costh{}(t)$.

\subsection{Presentation of Experimental Data}
Displaying the acceleration and rotation frequency of the molecules in Fig.~2 \& 4 is assisted by calculating the short time Fourier transform (STFT) of the measurement signal. During data acquisition, it was ensured that the sampling rate is greater than the Nyquist rate for the expected oscillation frequency. Because of varying measurement point density (notice Fig.~2 \textbf{(c)}) linear splines were used to resample onto a uniform grid. Further, the STFT was calculated using a $\SI{180}{\pico\second}$ wide blackman window.

\subsection{Simulation parameters}\label{parameters}
$H_{\rm rot}^{\rm RF}(t)$ is represented in the truncated $|J,M\rangle$ basis ($J\le24$, even $J$ only, 169 states) and diagonalized at each of 500 time steps spanning the Gaussian intensity envelope of the pulse, using the gas-phase constants $B = \SI{3.27}{\giga\hertz}$, $\Delta\alpha = \SI{7.77}{\angstrom^3}$ for \CSS{} and $B = \SI{6.08}{\giga\hertz}$, $\Delta\alpha = \SI{4.67}{\angstrom^3}$ for OCS, a droplet temperature $T=\SI{0.37}{\kelvin}$, and peak centrifuge intensity $I = \SI[parse-numbers = false]{3\cdot 10^{10}}{W/cm^2}$. The resulting eigenstates and eigenvalues are used to evaluate the steady-state density matrix of Eq.~(2). The thermalization time $\tau$, the only free parameter of the model, was fit to $\SI{10}{\pico\second}$ for \CSS{} and $\SI{100}{\pico\second}$ for OCS (see Theoretical Methods below for the full description of the model).

\subsubsection{Rescaling of \costh{} to account for non-axial recoil}
The simulated \costh{}$(t)$ was rescaled about the isotropic value of $1/2$ as
\begin{equation}
  \costh{}_{\rm scaled}(t) = \frac{1}{2} + \alpha\left(\costh{}(t) - \frac{1}{2}\right),
\end{equation}
with the amplitude factor $\alpha$ obtained from a least-squares fit to the experimental data, accounting for the non-axial recoil of fragments upon Coulomb explosion \cite{Christensen2016, Chatterley2020}: $\alpha = 0.069$ for \CSS{} and $\alpha = 0.55$ for OCS.

\section{Theoretical Methods}
We consider a linear rigid rotor of rotational constant $B$ driven by a strong optical centrifuge and coupled to a bosonic bath. The centrifuge rotation axis defines the space-fixed $z$ axis, and the laser polarization direction at the reference time defines the $x$ axis. Most of calculations are performed in the rotating frame at angular velocity $\Omega_0$, in which the drive has a static orientation.

The free rotor Hamiltonian is
\begin{equation}
H_{\rm 0} = B \mathbf{J}^2 ,
\end{equation}
with eigenstates $|J M\rangle$ satisfying
\begin{equation}
\mathbf{J}^2 |J M\rangle = J(J+1)|J M\rangle,
\qquad
J_z |J M\rangle = M |J M\rangle .
\end{equation}

The optical centrifuge produces an anisotropic potential aligned in the laboratory frame. In the rotating frame at angular velocity $\Omega_0$, the drive becomes time-independent and reads
\begin{equation}
H_{\rm drv} = -V_0 \sin^2\theta \cos^2\phi ,
\end{equation}
where $V_0 = \frac{\Delta\alpha}{4} E_0^2$ is set by the molecular polarizability anisotropy and the field amplitude.

The rotating-frame Hamiltonian is therefore
\begin{equation}
H_{\rm rot}^{\rm RF}
=B\mathbf{J}^2-\Omega_0J_z-V_0\sin^2\theta\cos^2\phi.
\label{eq:rotor_rf}
\end{equation}

The term $-\Omega_0 J_z$ lifts the degeneracy in $M$.

\subsection{Strong-Drive Regime and Localized Rotor}

In the strong-driving regime,
\begin{equation}
V_0 \gg B ,
\end{equation}
the centrifuge potential dominates over the bare rotational kinetic energy and the rotor becomes localized near the minimum of the drive potential.

The centrifuge potential
\begin{equation}
V_{\rm drv}(\theta,\phi) = -V_0 \sin^2\theta \cos^2\phi
\end{equation}
has minima at
\begin{equation}
\theta_0 = \frac{\pi}{2},
\qquad
\phi_0 = 0 \quad (\text{mod } \pi).
\end{equation}

We expand around the minimum $(\theta_0,\phi_0)$ and define small fluctuations
\begin{equation}
x = \delta\theta = \theta - \frac{\pi}{2},
\qquad
y = \delta\phi = \phi .
\end{equation}

Using
\begin{equation}
\sin\theta \simeq 1 - \frac{x^2}{2},
\qquad
\cos\phi \simeq 1 - \frac{y^2}{2},
\end{equation}
the potential becomes, to quadratic order,
\begin{equation}
V_{\rm drv}
\simeq
- V_0
+ V_0 (x^2 + y^2).
\end{equation}

Thus the minimum is harmonic and isotropic in the local tangent plane.

The rotor kinetic operator in angular coordinates is
\begin{equation}
H_{\rm 0} = B \mathbf{J}^2
= -B \left[
\frac{1}{\sin\theta}\partial_\theta
(\sin\theta\,\partial_\theta)
+ \frac{1}{\sin^2\theta}\partial_\phi^2
\right].
\end{equation}

Near $\theta=\pi/2$, one has $\sin\theta \simeq 1$, and the metric becomes locally flat:
\begin{equation}
ds^2 \simeq dx^2 + dy^2.
\end{equation}

To leading order,
\begin{equation}
H_{\rm 0}
\simeq
- B \left(
\partial_x^2 + \partial_y^2
\right).
\end{equation}

Using $I = \frac{1}{2B}$ (with $\hbar=1$), this can be written as
\begin{equation}
H_{\rm 0}
=
\frac{p_x^2 + p_y^2}{2I}.
\end{equation}

Combining kinetic and expanded potential terms, the Hamiltonian of the rotor in the rotating frame becomes
\begin{equation}
H_{\rm rot}^{\rm RF}
=
\frac{p_x^2 + p_y^2}{2I}
+ \frac12 I \Omega_{\rm loc}^2 (x^2 + y^2)
- \Omega_0 J_z
- V_0,
\end{equation}
with harmonic frequency
\begin{equation}
\Omega_{\rm loc} = 2\sqrt{B V_0}.
\end{equation}

In the local approximation,
\begin{equation}
J_z = -i\partial_\phi \simeq -i\partial_y = p_y.
\end{equation}

Therefore
\begin{equation}
H_{\rm rot}^{\rm RF}
=
\frac{p_x^2}{2I}
+
\frac{p_y^2}{2I}
+ \frac12 I\omega^2_h(x^2+y^2)
- \Omega_0 p_y
- V_0.
\end{equation}

Completing the square,
\begin{equation}
H_{\rm rot}^{\rm RF}
=
\frac{p_x^2}{2I}
+
\frac{(p_y - I\Omega_0)^2}{2I}
+
\frac12 I\omega^2_h(x^2+y^2)
-
\frac{I\Omega_0^2}{2}
- V_0.
\end{equation}

The Hamiltonian is unitarily equivalent to a standard two-dimensional isotropic harmonic oscillator. The eigenenergies are therefore
\begin{equation}
E_{n_x,n_y}(\Omega_0)
=
- V_0
+
\Omega_{\rm loc} (n_x+n_y+1)
-
\frac{I\Omega_0^2}{2},
\qquad
n_x,n_y \in \mathbb{N}.
\end{equation}

All levels are shifted by the same constant $-I\Omega_0^2/2$, reflecting the subtraction of rigid rotational kinetic energy in the rotating frame.

The eigenfunctions acquire a phase factor
\begin{equation}
\psi_{n_x,n_y}(\Omega_0)
=
e^{i I\Omega_0 y}
\,
\psi_{n_x,n_y}(0),
\end{equation}
while their spatial profile remains that of the two-dimensional harmonic oscillator centered at the potential minimum.

\subsection{Dominant Angular Momentum State and Localization Condition}

The same localization scale follows from direct minimization in the $|J,M\rangle$ basis.

When the Hamiltonian in Eq.~\eqref{eq:rotor_rf} confines the molecule near $(\theta_0,\phi_0) = (\pi/2,0)$, the drive contributes $\approx -V_0$. The dominant angular momentum follows by minimizing the rotational and Coriolis terms over $|J,M\rangle$ yielding $M^*=J$ and
\begin{equation}
\frac{d}{dJ}\left[BJ(J+1) - \Omega_0 J\right] = 0
\qquad\Longrightarrow\qquad
J^* \approx \frac{\Omega_0}{2B}.
\end{equation}
Thus, the molecule co-rotates at frequency $\Omega_0$, carrying angular momentum $J^* = \Omega_0/(2B)$.

The rotational kinetic energy of this dominant state is
\begin{equation}
E_{\rm kin} = B J^*(J^*+1) \approx \frac{\Omega_0^2}{4B}.
\end{equation}
For the drive to trap the molecule, this kinetic energy must remain smaller than the drive depth $V_0$:
\begin{equation}
E_{\rm kin} \ll V_0
\qquad\Longrightarrow\qquad
\Omega_0 \ll 2\sqrt{BV_0} = \Omega_{\rm loc}.
\end{equation}
Hence localization requires $\Omega_0\ll\Omega_{\rm loc}$.

\subsection{Matrix Representation and Numerical Diagonalization}

We represent the rotating-frame Hamiltonian of Eq.~\eqref{eq:rotor_rf} in the $|J,M\rangle$ basis, truncated at $J\le J_{\rm max}$.

The kinetic and Coriolis terms are diagonal:
\begin{equation}
\langle J,M|B\mathbf{J}^2 - \Omega_0 J_z|J',M'\rangle
=\left[BJ(J+1) - \Omega_0 M\right]\delta_{JJ'}\delta_{MM'}.
\end{equation}

For the drive term, we use the decomposition
\begin{equation}
\sin^2\theta\cos^2\phi
=
\frac{1}{3}
- \frac{1}{3}\sqrt{\frac{4\pi}{5}}\,Y_2^0
+\sqrt{\frac{2\pi}{15}}\left(Y_2^{+2}+Y_2^{-2}\right),
\end{equation}
so the drive couples only states with $\Delta J = 0,\pm 2$ and $\Delta M = 0, \pm 2$. Its matrix elements are given by Gaunt coefficients:
\begin{equation}
\langle J_a M_a | Y_{\lambda\mu} | J_b M_b \rangle
=
(-1)^{M_a}
\sqrt{\frac{(2J_a+1)(2\lambda+1)(2J_b+1)}{4\pi}}
\begin{pmatrix}
J_a & \lambda & J_b \\
0 & 0 & 0
\end{pmatrix}
\begin{pmatrix}
J_a & \lambda & J_b \\
- M_a & \mu & M_b
\end{pmatrix}.
\label{eq:sup_gaunt}
\end{equation}

Numerical diagonalization of the Hermitian matrix of $H_{\rm rot}^{\rm RF}$ yields eigenstates $|\tilde n\rangle$ and eigenenergies $\tilde E_n$,
\begin{equation}
H_{\rm rot}^{\rm RF}|\tilde n\rangle = \tilde E_n\,|\tilde n\rangle,
\qquad
|\tilde n\rangle = \sum_{J,M} c_{n,JM}\,|J,M\rangle.
\end{equation}
The ground state $|\tilde 0\rangle$ corresponds to the molecule localized near $(\theta_0,\phi_0) = (\pi/2,0)$ with angular momentum structure controlled by $\Omega_0$ and $V_0$.

\subsection{Shared Pulse and Detection Conventions}
\label{sec:sup_shared_conventions}

The $|J,M\rangle$ matrix representation of Eq.~\eqref{eq:rotor_rf} is used throughout. The pulse and detection conventions below are common to the gas-phase, steady-state, and dressed-Lindblad calculations.

The centrifuge frequency follows a linear chirp,
\begin{equation}
\Omega_0(t) = 2\pi\left[f_0 + {\rm ramp}\cdot t\right],
\end{equation}
and the laser intensity follows a Gaussian envelope $g(t)$ centered at $t=0$. The gas-phase and steady-state calculations use a FWHM of $320$ ps, whereas the Lindblad pulse width is fitted separately. The drive coupling is
\begin{equation}
V_0(t) = \frac{\Delta\alpha}{4}\,E^2(t) = V_0\, g(t),
\end{equation}
with $E^2(t) = E_0^2\, g(t)$.

The experiment measures the angle $\theta_{2D}$ between the molecular axis projected onto the detection plane (the plane perpendicular to the VMI axis) and the instantaneous centrifuge polarization direction. In the rotating frame,
\begin{equation}
\cos^2\theta_{2D}
=
\frac{\sin^2\theta\cos^2\phi}{\sin^2\theta\cos^2\phi + \cos^2\theta},
\end{equation}
where $\phi$ is measured from the instantaneous polarization axis. For a perfect alignment along the centrifuge, this observable equals 1 when the molecular axis is aligned with the detection plane ($\phi=0, \theta=\pi/2$) and $1/2$ it is perpendicular ($\phi=\pi/2, \theta=\pi/2$).

The observable matrix is precomputed in the $|J,M\rangle$ basis. For comparison with experiment, the calculated trace $S(t)=\langle\cos^2\theta_{2D}\rangle(t)$ is mapped through the display contrast,
\begin{equation}
S_\alpha(t) = \frac12 + \alpha\left(S(t) - \frac12\right),
\end{equation}
where $\alpha$ accounts for contrast not captured by the single-intensity calculation. The displayed OCS and CS$_2$ gas-phase traces use $\alpha=1$ and $0.5$, respectively.

\subsection{Gas-Phase Hamiltonian Dynamics}

The rotor model is compared against gas-phase measurements, in which no helium-bath terms are present. The full time-dependent Schr\"odinger equation is propagated in the standard $|J,M\rangle$ basis. The field-free ensemble is prepared at $T=10\;\mathrm{K}$ with Boltzmann weights $w_{JM}\propto e^{-\beta E_J}$ where every $M=-J,\ldots,J$ is included. OCS contains both even and odd $J$; for $^{12}$C$^{32}$S$_2$, nuclear-spin statistics restrict the ensemble to even $J$ but do not restrict $M$. The Boltzmann cutoff retains $J\leq12$ for OCS (169 initial states) and even $J\leq16$ for CS$_2$ (153 initial states). The propagation bases extend to $J_{\max}=20$ and 25, respectively. Both calculations cover $-600\leq t\leq350\;\mathrm{ps}$ and use relative and absolute ODE tolerances of $10^{-11}$. The slope of the acceleration is fitted to the data; the two accelerations therefore differ slightly between the molecules.

\begin{center}
\begin{tabular}{lcc}
 & OCS & CS$_2$ \\
\hline
$B/h$ (GHz) & 6.082 & 3.271 \\
$\Delta\alpha$ (\AA$^3$) & 4.67 & 9.60 \\
$I_0$ (W/cm$^2$) & $3.0\times10^{11}$ & $3.0\times10^{11}$ \\
$\tau_{\rm FWHM}$ (ps) & 320 & 320 \\
$f(0)$ (GHz) & 25.5 & 34.0 \\
$df/dt$ (MHz/ps) & 160 & 157 \\
$V_{\rm pk}/h$ (GHz) & 443 & 911 \\
$J_{\max}$ & 20 & 25 \\
display contrast $\alpha$ & 1 & 0.5 \\
\end{tabular}
\end{center}

\begin{figure}[p]
    \centering
    \includegraphics[width=0.98\textwidth]{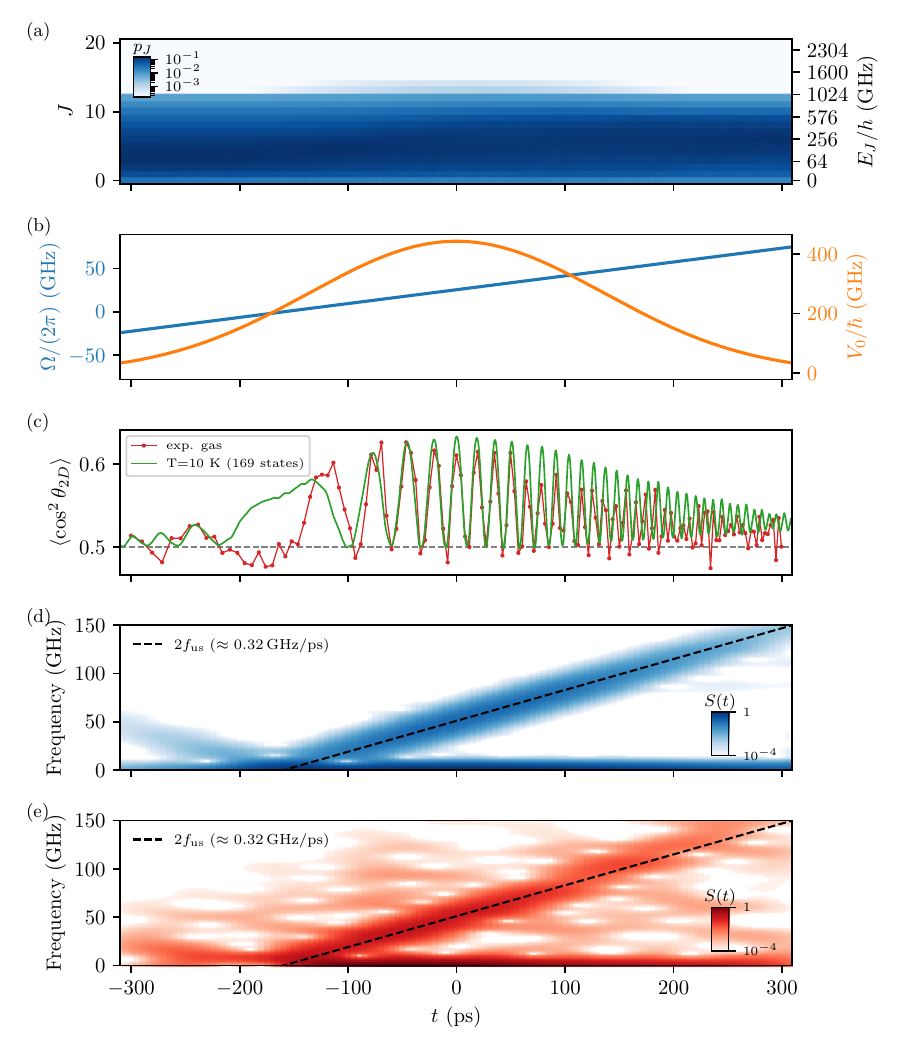}
    \caption{\textbf{Helium-free OCS benchmark.} Pulse parameters: $I_0=3.0\times10^{11}\;\mathrm{W/cm^2}$, $\tau_{\rm FWHM}=320\;\mathrm{ps}$, $f(0)=25.5\;\mathrm{GHz}$, and $df/dt=160\;\mathrm{MHz/ps}$. The $T=10\;\mathrm{K}$ calculation averages 169 initial $|J,M\rangle$ states and uses $J_{\max}=20$.
    \textbf{(a)} Thermally averaged $J$-state populations (logarithmic color scale).
    \textbf{(b)} Centrifuge frequency $\Omega(t)/2\pi$ (blue) and laser coupling $V_0(t)/\hbar$ (orange), both in GHz.
    \textbf{(c)} Laboratory-frame $\langle\cos^2\theta_{\mathrm{2D}}\rangle$ with experimental gas data (orange markers) and the unscaled thermal calculation (green).
    \textbf{(d)} Gaussian-windowed moving Fourier power $|\mathrm{STFT}|^2$ of the calculated alignment trace ($\sigma_t=30\;\mathrm{ps}$).
    \textbf{(e)} Same STFT applied to the experimental gas data, with $2f_{\rm us}(t)$ overlay (dashed).}
    \label{fig:sup_fig1}
\end{figure}

\begin{figure}[p]
    \centering
    \includegraphics[width=0.88\textwidth]{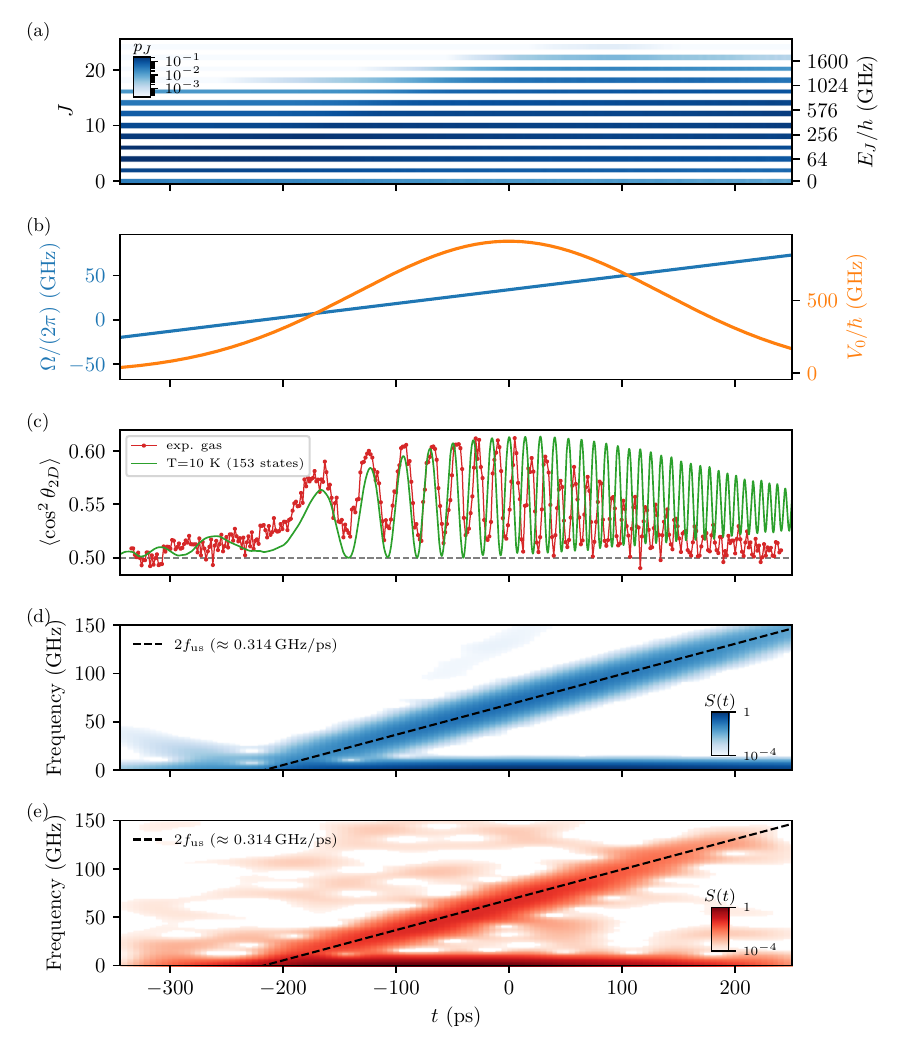}
    \caption{\textbf{Helium-free CS$_2$ benchmark.} Pulse parameters: $I_0=3.0\times10^{11}\;\mathrm{W/cm^2}$, $\tau_{\rm FWHM}=320\;\mathrm{ps}$, $f(0)=34.0\;\mathrm{GHz}$, and $df/dt=157\;\mathrm{MHz/ps}$. The $T=10\;\mathrm{K}$ calculation averages 153 even-$J$, all-$M$ initial states and uses $J_{\max}=25$.
    \textbf{(a)} Thermally averaged $J$-state populations at $T=10\;\mathrm{K}$ (log colors).
    \textbf{(b)} Centrifuge frequency $\Omega(t)/2\pi$ (blue) and laser coupling $V_0(t)/\hbar$ (orange), both in GHz.
    \textbf{(c)} Laboratory-frame $\langle\cos^2\theta_{\mathrm{2D}}\rangle$ with experimental gas data (orange markers). The green thermal trace is displayed as $1/2+0.5[\langle\cos^2\theta_{\mathrm{2D}}\rangle-1/2]$.
    \textbf{(d)} Gaussian-windowed moving Fourier power $|\mathrm{STFT}|^2$ of the thermal alignment trace.
    \textbf{(e)} Same STFT applied to the experimental gas data, with $2f_{\rm us}(t)$ overlay (dashed).}
    \label{fig:sup_fig2}
\end{figure}

The stronger CS$_2$ coupling in Figure~\ref{fig:sup_fig2} produces a broader, higher-$J$ wavepacket than for OCS in Figure~\ref{fig:sup_fig1}. For CS$_2$, only even $J$ occur, while every allowed $M$ contributes to the thermal average. In both molecules, the calculated and measured STFTs follow the same $2f_{\rm us}(t)$ ridge. The CS$_2$ calculation, even when displayed with $\alpha=0.5$, retains excess late-time coherence, indicating decoherence absent from the Hamiltonian model.

\clearpage

\section{Renormalization by the Bosonic Bath}

The bath consists of bosonic excitations labelled by radial momentum $k$ and angular momentum quantum numbers $(\lambda,\mu)$,
\begin{equation}
H_{\rm bath}
=
\sum_{k\lambda\mu}
\omega_k \,
b_{k\lambda\mu}^\dagger b_{k\lambda\mu}.
\end{equation}

In the rotating frame, the bath dispersion is shifted according to the projection of angular momentum along the rotation axis,
\begin{equation}
H_{\rm bath}^{\rm RF}
=
\sum_{k\lambda\mu}
\left(\omega_k - \Omega_0 \mu\right)
b_{k\lambda\mu}^\dagger b_{k\lambda\mu}.
\end{equation}

This shift ensures total angular momentum conservation in the rotating frame.

We consider a linear coupling between the rotor orientation and bath modes,
\begin{equation}
H_{\rm int}
=
\sum_{k\lambda\mu}
U_\lambda(k)
\left[
Y_{\lambda\mu}(\hat\Omega)\,
b_{k\lambda\mu}
+
Y_{\lambda\mu}^\dagger(\hat\Omega)\,
b_{k\lambda\mu}^\dagger
\right],
\end{equation}
where $Y_{\lambda\mu}(\hat\Omega)$ are spherical harmonics acting on the rotor angular degrees of freedom.

The spherical-harmonic matrix elements are the Gaunt coefficients of Eq.~\eqref{eq:sup_gaunt}. The interaction preserves total angular momentum in the rotating frame.

Collecting all contributions, the complete Hamiltonian in the rotating frame is
\begin{equation}
H_{\rm tot}^{\rm RF}
=
B\mathbf{J}^2
- \Omega_0 J_z
- V_0 \sin^2\theta \cos^2\phi
+
\sum_{k\lambda\mu}
(\omega_k - \Omega_0 \mu)
b_{k\lambda\mu}^\dagger b_{k\lambda\mu}
+
H_{\rm int}.
\end{equation}

In the eigenbasis of $H_{\rm rot}^{\rm RF}$, the full Hamiltonian reads
\begin{equation}
H' = \sum_n \tilde E_n\,|\tilde n\rangle\langle\tilde n|
+\sum_{k\lambda\mu}(\omega_k - \Omega_0\mu)\,b^\dagger_{k\lambda\mu}b_{k\lambda\mu}
+ H_{\rm int},
\end{equation}
where
\begin{equation}
H_{\rm int}
=
\sum_{k\lambda\mu}\sum_{m,n}
U_\lambda(k)\,V^{\lambda\mu}_{mn}\,
|\tilde m\rangle\langle\tilde n|\,
b_{k\lambda\mu}
+ \text{h.c.}
\end{equation}
The coupling matrix elements are
\begin{equation}
V^{\lambda\mu}_{mn}
=
\langle\tilde m|Y_{\lambda\mu}|\tilde n\rangle
=
\sum_{J,M,J',M'}
c^*_{m,JM}\,c_{n,J'M'}\,
\langle J,M|Y_{\lambda\mu}|J',M'\rangle,
\end{equation}
with Gaunt coefficients on the right-hand side.

To second order in the rotor--bath coupling, the self-energy of eigenstate $|\tilde n\rangle$ is
\begin{equation}
\Sigma_n(\omega)
=
\sum_{k\lambda\mu}|U_\lambda(k)|^2
\sum_m\frac{|V^{\lambda\mu}_{mn}|^2}
{\omega - \tilde E_m - \omega_k + \Omega_0\mu + i\eta}.
\label{eq:sup_driven_selfenergy}
\end{equation}
The imaginary part evaluated on-shell gives the damping rate,
\begin{equation}
\Gamma_n = -2\,\operatorname{Im}\,\Sigma_n(\tilde E_n).
\end{equation}

This self energy describes the energy renormalisation and relaxation rate within the weak coupling approximation.

In practice however, we find numerically $\re{\Sigma_n}\sim$ GHz, small compared to the typical energy spicing of the driven rotor $\hbar \Omega_{\rm loc}/2\pi\sim 30$ GHz. Therefore, the direct renormalisation of the modes of the driven rotor only affects weakly the dynamics and can be neglected. Note in particular that the bath does not generate any centrifugal wall for the driven rotor.

Thus, the bath rather acts as a slow relaxation mechanism, and the molecular dynamics is primarily governed by $H_{\rm rot}^{\rm RF}$. This supports treating the bath coupling perturbatively and justifies the thermalization picture developed in next Section.

\section{Thermalization of a Rotating Driven Rotor}
\label{sec:sup_thermalisation}

We consider a quantum rotor with Hamiltonian
\begin{equation}
H_{\rm rot}^{\rm LF}(t) = H_0 + R^\dagger(t)\, V_0 \, R(t),
\end{equation}
in the lab frame,
where
\begin{equation}
H_0 = B J^2,
\end{equation}
and $R(t)=e^{-i\Omega t J_z}$ describes a rigid rotation of the potential $V_0$ around the $z$-axis with angular velocity $\Omega$.
Going to the rotating frame, the Hamiltonian becomes time-independent:
\begin{equation}
H_{\mathrm{rot}}^{\rm RF} = H_0 + V_0 - \Omega J_z.
\end{equation}

The laboratory-frame density matrix transforms as
\begin{equation}
\rho^{\rm RF}(t) = R(t)\,\rho^{\rm LF}(t)\, R^\dagger(t).
\end{equation}

The system is coupled to a thermal bath at inverse temperature $\beta$, with phenomenological relaxation time $\tau$. The corresponding instantaneous laboratory-frame Gibbs state is
\begin{equation}
\rho_{\mathrm{th}}^{\mathrm{LF}}(t) = \frac{e^{-\beta H_{\rm rot}^{\rm LF}(t) }}{Z(t)}.
\end{equation}
In the rotating frame this state is time independent:
\begin{equation}
\rho_{\mathrm{th}}^{\mathrm{RF}} = \frac{e^{-\beta(H_0 + V_0)}}{Z}.
\end{equation}
This is not the thermal state obtained from the Hamiltonian in the rotating frame $H_{\mathrm{rot}}^{\rm RF}$,
\begin{equation}
\rho_{\mathrm{th}}^{\mathrm{RF}} \neq \frac{e^{-\beta H_{\mathrm{rot}}^{\rm RF}}}{Z}.
\end{equation}
because the latter includes the acceleration term. A bath at equilibrium in the laboratory frame relaxes the rotor toward $\rho_{\mathrm{th}}^{\mathrm{LF}}(t)$, or equivalently $\rho_{\mathrm{th}}^{\mathrm{RF}}$, rather than toward the Gibbs state of $H_{\mathrm{rot}}^{\rm RF}$.

To model this, we adopt a phenomenological master equation in the rotating frame:
\begin{equation}
\partial_t \rho^{\rm RF} = -i[H_{\mathrm{rot}}^{\rm RF} , \rho^{\rm RF} ] - \frac{1}{\tau}(\rho^{\rm RF} - \rho_{\mathrm{th}}^{\rm RF}),
\end{equation}

At steady state,
\begin{equation}
0 = -i[H_{\mathrm{rot}}^{\rm RF} , \rho_{\mathrm{ss}}^{\rm RF} ] - \frac{1}{\tau}(\rho_{\mathrm{ss}}^{\rm RF}  - \rho_{\mathrm{th}}^{\mathrm{RF}}).
\end{equation}
Let $\{|n\rangle\}$ be eigenstates of $H_{\mathrm{rot}}^{\rm RF} $:
\begin{equation}
H_{\mathrm{rot}}^{\rm RF}  |n\rangle = E_n |n\rangle.
\end{equation}
In this basis, the solution is
\begin{equation}
(\rho_{\mathrm{ss}}^{\rm RF} )_{nm}
=
\frac{(\rho_{\mathrm{th}}^{\mathrm{RF}})_{nm}}{1 + i\tau (E_n - E_m)}.
\end{equation}

In the fast-relaxation limit $\tau \to 0$,
\begin{equation}
\rho_{\mathrm{ss}}^{\rm RF}  \to \rho_{\mathrm{th}}^{\mathrm{RF}} = \frac{e^{-\beta(H_0+V_0)}}{Z}.
\end{equation}
In the slow-relaxation limit $\tau \to \infty$, off-diagonal terms vanish:
\begin{equation}
(\rho_{\mathrm{ss}}^{\rm RF} )_{nm} \to 0 \quad (n \neq m),
\end{equation}
while diagonal terms remain:
\begin{equation}
(\rho_{\mathrm{ss}}^{\rm RF} )_{nn} = (\rho_{\mathrm{th}}^{\mathrm{RF}})_{nn}.
\end{equation}
Thus,
\begin{equation}
\rho_{\mathrm{ss}}^{\rm RF}  \approx \sum_n |n\rangle\langle n| \, \langle n|\rho_{\mathrm{th}}^{\mathrm{RF}}|n\rangle.
\end{equation}
The steady state is the projection of $\rho_{\mathrm{th}}^{\mathrm{RF}}$ onto operators commuting with $H_{\mathrm{rot}}^{\rm RF} $.

\section{Instantaneous Steady-State Calculation}
\label{sec:sup_steady_state_numerics}

The instantaneous steady-state calculation evaluates, at each time, the frozen attractor of the master equation of Section~\ref{sec:sup_thermalisation}. We note that its relaxation parameter $\tau$ is a phenomenological parameter that is qualitatively different from the three Lindblad rates $\gamma_{\rm inel}$, $\gamma_{\rm el}$, $\gamma_\phi$ which are used in the dressed-Lindblad calculation later.

\subsection{Parameters and Basis}

The helium bath temperature is $T = 0.37$ K.
The drive coupling strength is $V_0 = \frac{1}{4}\Delta\alpha\, E_0^2$, with laser amplitude $I = \frac{1}{2}c\epsilon_0 E_0^2 =3\cdot 10^{10}$ W/cm$^2$ at its peak. We can also rewrite $V_0 = \frac{h}{2}\Delta\alpha_v\, I_v$ where $\Delta\alpha_v=\Delta\alpha/4\pi\epsilon_0$ is the polarizability volume and $I_v=2 I/\hbar c$ with a peak amplitude of $I_v = 19$ GHz.\AA$^3$. The Gaussian intensity FWHM is $320$ ps; the chirp follows Section~\ref{sec:sup_shared_conventions}.

Both calculations use $J_{\rm max}=25$. Because the centrifuge potential $\sin^2\theta\cos^2\phi$ is even under the antipodal map $(\theta,\phi)\to(\pi-\theta,\phi+\pi)$, it connects only states of the same parity $(-1)^J$. OCS includes all $J$ and has 676 basis states. For CS$_2$, nuclear-spin statistics retain even $J$ through 24, giving 325 states.

For CS$_2$ and OCS we use the following parameters:
\begin{center}
\begin{tabular}{lcc|cc}
Physical parameter & Notation & Unit  & CS$_2$ & OCS \\
\hline
Rotational constant & $B$ & GHz & 3.27 & 6.08 \\
Polarizability anisotropy & $\Delta\alpha$  & \AA$^3$ & 7.77 & 4.67 \\
Maximal pendular frequency $\sqrt{B\Delta\alpha E_0^2}$ & $\Omega_{\rm loc}/2\pi$ & GHz & 31  & 33\\
Interaction energy with He & & meV & 7 & 3.5 \\
Interaction energy anisotropy with He & & meV & 4.8 & 1.6 \\
Relaxation time & $\tau$ & ns & 0.01 & 0.1 \\
Typical relaxation dephasing & $\tau\Omega_{\rm loc}/2\pi$ & 1 & 2 & 20 \\
Display contrast & $\alpha$ & 1 & 0.069 & 0.546 \\
\end{tabular}
\end{center}
The optical centrifuge parameters (initial frequency $f_0$, linear chirp rate, initial phase $\phi_0$) are fitted from the experimental oscillation envelope:
\begin{center}
\begin{tabular}{lccc}
 & $f_0$ (GHz) & ramp (MHz/ps) & $\phi_0$ (rad) \\
\hline
CS$_2$ & 18.997 & 92.08 & 0.0110 \\
OCS & 22.499 & 97.59 & 0.5720 \\
\end{tabular}
\end{center}

\subsection{Steady-State Construction}

At each of $N_t = 500$ uniformly spaced time steps, the rotating-frame Hamiltonian of Eq.~\eqref{eq:rotor_rf} is assembled in the $|J,M\rangle$ basis with the time-dependent frequency and coupling of Section~\ref{sec:sup_shared_conventions} and the bare rotational constant $B$, and diagonalized by full dense diagonalization, yielding instantaneous eigenstates $\{|\tilde n(t)\rangle, \tilde E_n(t)\}$. The instantaneous steady state is then constructed as follows:
\begin{enumerate}
\item Compute the target Boltzmann density matrix of the non-rotating Hamiltonian
\begin{equation}
H_{\rm target}(t) = B\mathbf{J}^2 - V_0(t)\sin^2\theta\cos^2\phi
\end{equation}
at temperature $T = 0.37$ K:
\begin{equation}
\rho_{\mathrm{th}}^{\mathrm{RF}}(t) = \frac{e^{-\beta H_{\rm target}(t)}}{Z(t)}.
\end{equation}
\item Apply the steady-state formula of Section~\ref{sec:sup_thermalisation} in the instantaneous eigenbasis:
\begin{equation}
\left(\rho_{\rm ss}^{\mathrm{RF}}(t)\right)_{mn}
=
\frac{\left(\rho_{\mathrm{th}}^{\mathrm{RF}}(t)\right)_{mn}}{1 + i\tau\left(\tilde E_m(t) - \tilde E_n(t)\right)},
\end{equation}
where $\tau$ is the bath relaxation timescale, fitted to $\tau = 10$ ps for CS$_2$ and $\tau = 100$ ps for OCS.
\end{enumerate}
Observables are evaluated as $\langle O\rangle(t) = \operatorname{Tr}[\rho_{\rm ss}^{\mathrm{RF}}(t)\,O(t)]$, using the 2D detector observable and display contrast of Section~\ref{sec:sup_shared_conventions}. Because the density matrix is not propagated in time, this quasistatic construction does not describe transient lag or memory of the initial state.

\begin{figure}
\centering
\includegraphics[width=\textwidth]{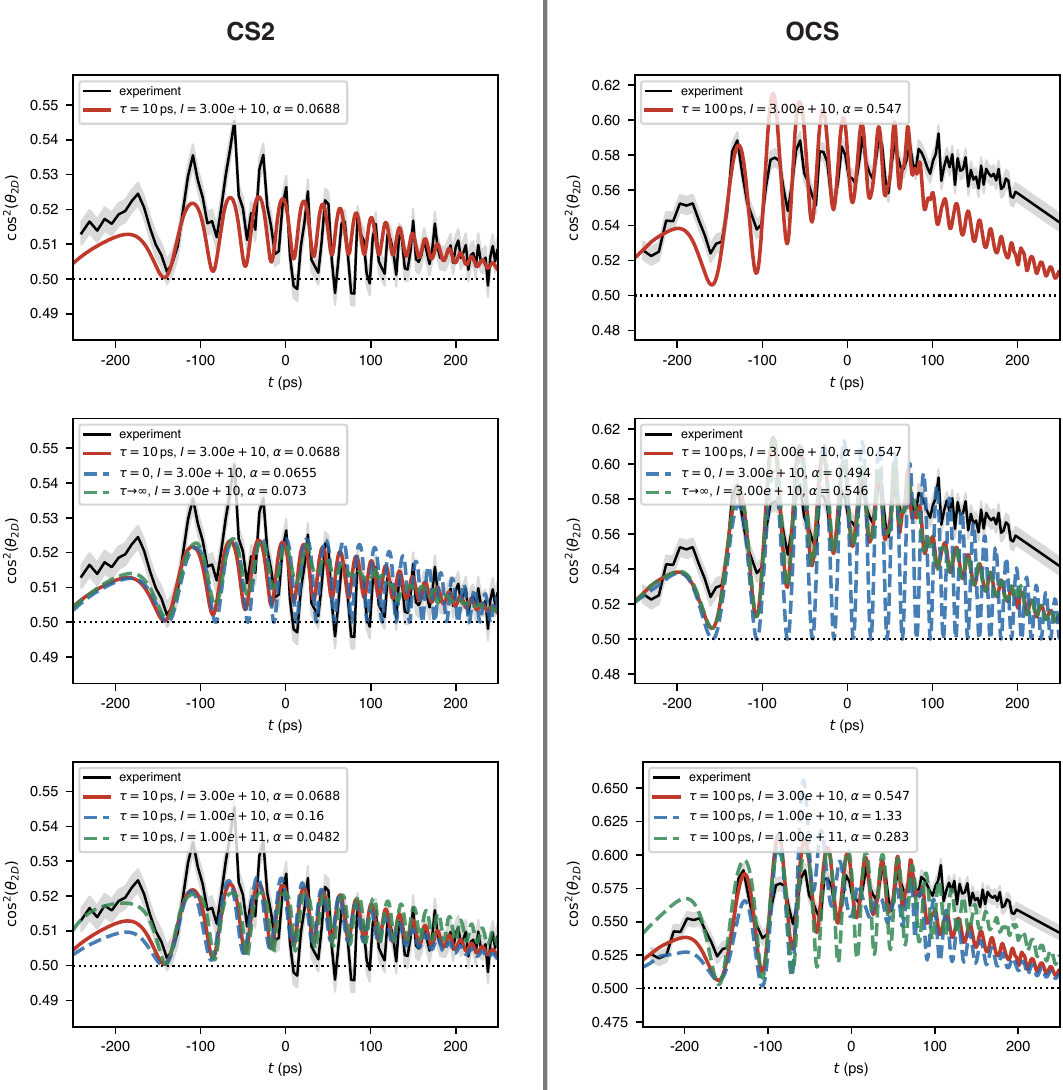}
\caption{\textbf{Instantaneous steady-state parameter scan.} Columns show CS$_2$ (left) and OCS (right). Rows compare the selected model with experiment, vary the relaxation time $\tau$, and vary the peak intensity $I$, respectively.}
\label{fig:sup_steady_state_scan}
\end{figure}

\clearpage

\section{Dressed-Lindblad Dynamics}
\label{sec:sup_dressed_lindblad}

The time-dependent Lindblad propagation keeps the transient dynamics omitted by the steady-state construction~\cite{Gorini1976,Lindblad1976}.

\begin{figure}[p]
    \centering
    \includegraphics[width=0.88\textwidth]{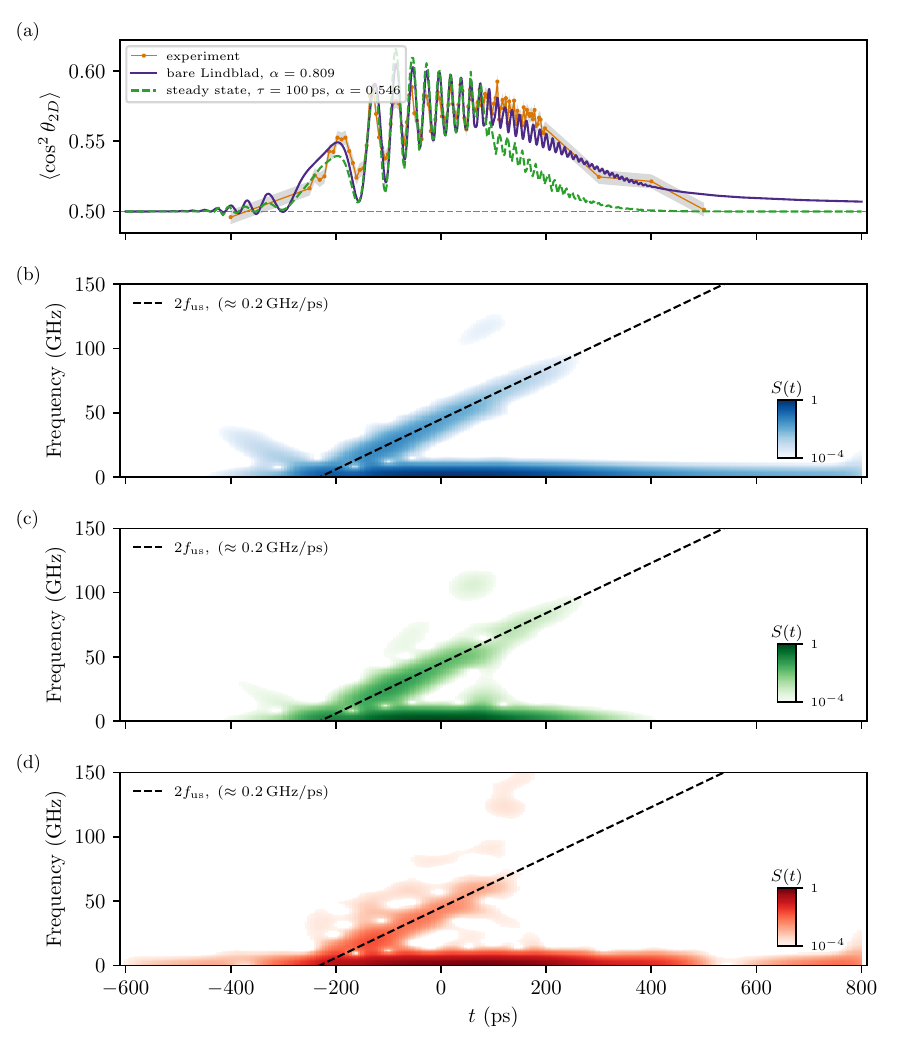}
    \caption{\textbf{Dressed-Lindblad evolution for OCS} using the parameters in Table~\ref{tab:sup_lindblad_parameters}.
    \textbf{(a)} Alignment signal: propagated dressed-Lindblad model (purple, $\alpha=0.888$), instantaneous steady state (green dashed, $\tau=100\;\mathrm{ps}$ and $\alpha=0.546$), and experiment (orange).
    \textbf{(b)} STFT of the dressed-Lindblad signal.
    \textbf{(c)} STFT of the steady-state signal.
    \textbf{(d)} Experimental STFT. The dashed lines indicate $2f_{\rm us}(t)$.}
    \label{fig:sup_dressed_lindblad_ocs}
\end{figure}

\begin{figure}[p]
    \centering
    \includegraphics[width=0.88\textwidth]{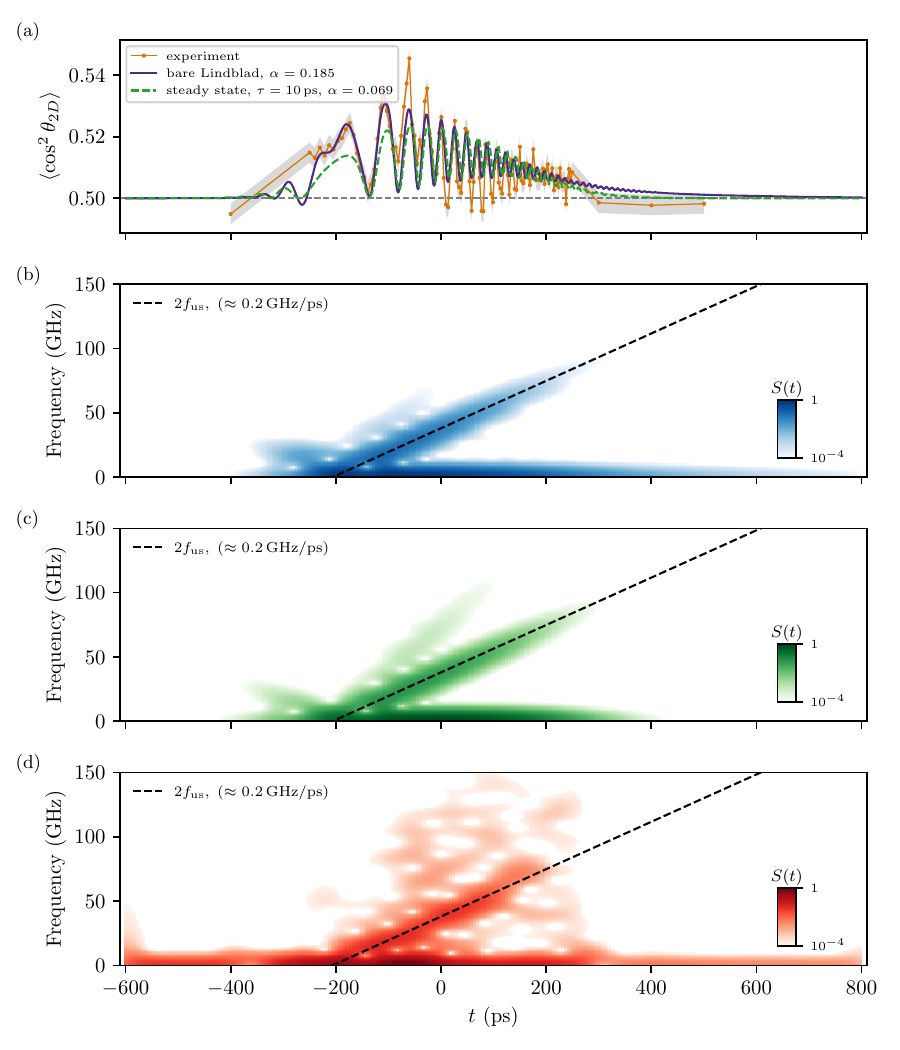}
    \caption{\textbf{Dressed-Lindblad evolution for CS$_2$} using the parameters in Table~\ref{tab:sup_lindblad_parameters}.
    \textbf{(a)} Alignment signal: propagated dressed-Lindblad model (purple, $\alpha=0.371$), instantaneous steady state (green dashed, $\tau=10\;\mathrm{ps}$ and $\alpha=0.069$), and experiment (orange).
    \textbf{(b)} STFT of the dressed-Lindblad signal.
    \textbf{(c)} STFT of the steady-state signal.
    \textbf{(d)} Experimental STFT. The dashed lines indicate $2f_{\rm us}(t)$.}
    \label{fig:sup_dressed_lindblad_cs2}
\end{figure}

The rotating-frame density matrix obeys
\begin{equation}
\begin{aligned}
\dot\rho={}&-i[H_{\rm RF}(t),\rho]
+\gamma_{\rm inel}\sum_{J\geq2}\sum_{\mu=-2}^{2}
 \left\{\mathcal D[A^-_{J\mu}]\rho
 +e^{-\beta\Delta E_J}\mathcal D[A^+_{J\mu}]\rho\right\} \\
&+\gamma_{\rm el}\sum_{J\geq1}\sum_{\mu=-2}^{2}\mathcal D[B_{J\mu}]\rho
+\gamma_\phi\sum_J\mathcal D[P_J]\rho,
\end{aligned}
\label{eq:sup_lindblad}
\end{equation}
with
\begin{equation}
H_{\rm RF}(t)=\sum_J E_JP_J-\Omega(t)J_z
-\kappa(t)\sin^2\theta\cos^2\phi,
\qquad
\mathcal D[L]\rho=L\rho L^\dagger-\tfrac12\{L^\dagger L,\rho\}.
\label{eq:sup_lindblad_hamiltonian}
\end{equation}
The normalized rank-2 operators are defined by
\begin{equation}
\begin{aligned}
F^-_{J\mu}&=P_{J-2}Y_{2\mu}P_J,
&c_J^-&=\frac{1}{2J+1}\operatorname{Tr}\!\sum_{\mu=-2}^{2}
F_{J\mu}^{-\dagger}F^-_{J\mu},\\
F^0_{J\mu}&=P_JY_{2\mu}P_J,
&c_J^0&=\frac{1}{2J+1}\operatorname{Tr}\!\sum_{\mu=-2}^{2}
F_{J\mu}^{0\dagger}F^0_{J\mu},\\
A^-_{J\mu}&=F^-_{J\mu}/\sqrt{c_J^-},
&A^+_{J\mu}&=A_{J\mu}^{-\dagger},\qquad
B_{J\mu}=F^0_{J\mu}/\sqrt{c_J^0}.
\end{aligned}
\label{eq:sup_jump_operators}
\end{equation}
The dissipator follows the standard construction of covariant Markovian generators from irreducible tensor operators and the angular-momentum decomposition of rotor--bath coupling~\cite{Holevo1996,SchmidtLemeshko2015}. For a linear rotor in an isotropic helium environment, we keep the leading anisotropic rank-two contribution; hence, its spherical components $Y_{2\mu}$ generate the parity-preserving channels $\Delta J=0,\pm2$, and summing all five values of $\mu$ makes each dissipative channel rotationally covariant. The normalized operators $A^-_{J\mu}$ and $A^+_{J\mu}$ describe energy-changing relaxation, with the Boltzmann factor in Eq.~\eqref{eq:sup_lindblad} imposing the thermal bias between downward and upward transitions, while $B_{J\mu}$ describes reorientation within a $J$ shell and $P_J=\sum_M|JM\rangle\langle JM|$ dephases different $J-$shells. Henceforth, with non-negative rates, this GKSL form preserves trace and complete positivity~\cite{Gorini1976,Lindblad1976}. We therefore use it as a minimal phenomenological Lindblad ansatz consistent with rotational symmetry and thermal relaxation in helium~\cite{StienkemeierLehmann2006}; a microscopic derivation of the generator and its rates we postpone to a separate paper.
The initial state is the field-free Gibbs density at $T=0.37\;\mathrm K$, constructed from the Hamiltonian of Eq.~\eqref{eq:rotor_rf}. Parameter convergence is checked at $J_{\max}=12$ for OCS ($N=169$, all $J$) and $J_{\max}=14$ for CS$_2$ ($N=120$, even $J$ only), using adaptive Tsit5 with relative tolerance $10^{-9}$ and absolute tolerance $10^{-11}$. The smooth traces in Figures~\ref{fig:sup_dressed_lindblad_ocs} and~\ref{fig:sup_dressed_lindblad_cs2} are then recomputed on a $1\;\mathrm{ps}$ output grid at $J_{\max}=10$ and 14, respectively, with tolerances $10^{-7}$ and $10^{-9}$. The field amplitude is fixed at its independently estimated value, $e_0=1$ ($I_0=3\times10^{10}\;\mathrm{W/cm^2}$); only the three rates, pulse width, and profiled display contrast are fitted.

\begin{table}[htbp]
\centering
\caption{Parameters used in Figures~\ref{fig:sup_dressed_lindblad_ocs} and~\ref{fig:sup_dressed_lindblad_cs2}. Rates are in ps$^{-1}$ and times in ps.}
\label{tab:sup_lindblad_parameters}
\begin{tabular}{lcc}
\hline
 & OCS & CS$_2$ \\
\hline
$B^*$ (cm$^{-1}$) & 0.0731 & 0.02435 \\
$D^*$ (cm$^{-1}$) & $3.8\times10^{-4}$ & $4.0\times10^{-5}$ \\
$B$ (cm$^{-1}$) & 0.2029 & 0.1091 \\
$D$ (cm$^{-1}$) & $5.6\times10^{-8}$ & $1.2\times10^{-8}$ \\
$f(0)$ (GHz) & 22.499 & 18.997 \\
$df/dt$ (MHz/ps) & 97.590 & 92.083 \\
$\phi_0$ (rad) & 0.5720 & 0.0110 \\
$\sigma$ (ps) & 140 & 117 \\
$\gamma_{\rm inel}$ ($\tau_{\rm inel}$) & 0.010 (100) & 0.0005 (2000) \\
$\gamma_{\rm el}$ ($\tau_{\rm el}$) & 0.001 (1000) & 0.0193 (50) \\
$\gamma_\phi$ ($\tau_\phi$) & 0.020 (50) & 0.0123 (80) \\
display contrast $\alpha$ & 0.888 & 0.371 \\
$J_{\max}$ for convergence (basis size) & 12 (169) & 14 (120) \\
$J_{\max}$ in plotted trace (basis size) & 10 (121) & 14 (120) \\
maximum $J_\mathrm{max}$ population & $1.7\times10^{-9}$ & $2.1\times10^{-5}$ \\
\hline
\end{tabular}
\end{table}

\clearpage

For OCS, Figure~\ref{fig:sup_dressed_lindblad_ocs} shows that the steady-state curve follows the onset and oscillation frequency but returns to the isotropic baseline earlier than the propagated model. For CS$_2$, Figure~\ref{fig:sup_dressed_lindblad_cs2} shows the same principal driven ridge in both calculations, with a better transient amplitude and decay under explicit propagation. These differences isolate the lag and post-pulse memory absent from the quasistatic construction.

\end{document}